\iffalse
 \documentclass[Afour,sageh,times]{sagej}
 
 \newcommand{\mykeywords}[1]{\keywords{#1}}

 \newcommand{\myrunninghead}[1]{\runninghead{#1}}
 \newcommand{\mycorrauth}[1]{\corrauth{#1}}
 \newcommand{\myemail}[1]{\email{#1}}
 \newcommand{\pubmaketitle}{\maketitle}
 
 \else
 \documentclass[notitlepage]{article}
  \usepackage{xcolor}
  \usepackage[a4paper, total={6in, 8in}]{geometry}
  
  \newcommand{\mykeywords}[1]{\emph{Keywords}: #1}

  \newcommand{\mycorrauth}[1]{}
  \newcommand{\myemail}[1]{}
  \newcommand{\myrunninghead}[1]{}
   \newcommand{\pubmaketitle}{}
 
 \fi

\usepackage[utf8]{inputenc}

\usepackage{moreverb,url}

\ifpdf
\hypersetup{
  pdftitle={Portable High-order Finite Element Operator Kernels},
  pdfauthor={Noel Chalmers, and T. Warburton}
}
\fi

\usepackage{xcolor} 

\definecolor{mygreen}{rgb}{0,0.6,0}
\definecolor{mygray}{gray}{0.35}
\definecolor{mymauve}{rgb}{0.58,0,0.82}
\definecolor{mycodebg}{rgb}{1,0.96,0.83}
\definecolor{mytermbg}{rgb}{0.79,0.99,0.96}
\definecolor{mypythonbg}{rgb}{0.95,0.95,0.97}
\definecolor{myoutputbg}{rgb}{1.0,0.6, 0}
\definecolor{myexercise}{rgb}{0.52,0.80,0.98}
\definecolor{mymatlabbg}{rgb}{0.85,0.88,0.96}

\usepackage{listings}

\usepackage{algorithm}
\usepackage{algorithmic}
\usepackage{boxedminipage}

\usepackage{pgf,tikz}
\usepackage{pgfplots}
\usepackage{amsmath}
\usepackage{amsfonts}
\usepackage{subcaption}

\newcommand{\modelplot}[3]{
    \addplot [domain=0:1, mark=none,#1, thick, dashed ]
            { x/(#3 + x/#2)};}

\newcommand{\TM}{\textsuperscript{\texttrademark}}

\begin{document}

\title{Portable high-order finite element kernels I: Streaming Operations}
\author{Noel Chalmers\thanks{ AMD Research, Advanced Micro Devices Inc., 7171 Southwest Pkwy, Austin, TX, 78735. } \and Tim Warburton\thanks{Department of Mathematics, 225 Stanger Street, Virginia Tech, VA 24061.}}

\myrunninghead{Chalmers and Warburton}




\mycorrauth{Noel Chalmers}

\myemail{noel.chalmers@amd.com}

\maketitle

\begin{abstract}
    This paper is devoted to the development of highly efficient kernels performing vector operations relevant in linear system solvers. In particular, we focus on the low arithmetic intensity operations (i.e., streaming operations) performed within the conjugate gradient iterative method, using the parameters specified in the CEED benchmark problems \cite{ceed2017} for high-order hexahedral finite elements. We propose a suite of new Benchmark Streaming tests to focus on the distinct streaming operations which must be performed. We implemented these new tests using the OCCA abstraction framework \cite{medina2014occa} to demonstrate portability of these streaming operations on different GPU architectures, and propose a simple performance model for such kernels which can accurately capture data movement rates as well as kernel launch costs.
\end{abstract}

\mykeywords{High-order, finite element, GPU, AMD HIP, CUDA}
\pubmaketitle



\section{Introduction}


The current outlook for scientific applications in high performance computing (HPC) is a heterogeneous one, where high performance systems consist of a single or dual socket multi-core CPU paired with several GPU accelerator devices. This landscape has become the de facto norm since the introduction of the Summit and Sierra supercomputers, both of which utilize Nvidia V100 GPU accelerators. In these systems, effective use of the accelerators on each node is required to harness the system's high bandwidth and floating-point operation (FLOP) rates. Currently, on the horizon for HPC is the Aurora supercomputer which will use Intel-manufactured GPU accelerators, and the Frontier and El Capitan supercomputers which will use AMD GPU accelerators. There is therefore a need for HPC applications currently implemented with Nvidia's CUDA programming model to be portable to other device architectures, and to understand performance characteristics of these new platforms as we approach exascale computing.

A recently formed co-design center within the U.S. Department of Energy (DOE) Exascale Computing Project (ECP) is the Center for Efficient Exascale Discretizations. One of this co-design center's research thrusts has been to propose a suite of Benchmark Problems (BPs) \cite{ceed2017} to serve as performance tests for comparison of several software implementations of high-order finite element methods. Each benchmark problem consists of measuring the performance of solving a linear system with a matrix originating from different high-order finite element operators, via a Conjugate Gradient (CG) iterative method. In \cite{fischer2020scalability} the benchmark problems are discussed in detail, along with the stand-along operator Benchmark Kernel (BK) tests, originally proposed and studied in \cite{swirydowicz2019acceleration} on Nvidia P100 accelerators. Results from these studies have already been used in modernizing earlier benchmarks such as NekBone \cite{karp2020optimization}.

In this paper, we also consider the CEED BPs, but as opposed to \cite{swirydowicz2019acceleration} which focused on the FLOP-intensive finite element operator application, we instead focus on the necessary low arithmetic intensity vector operations required in the Conjugate Gradient iteration. Such low arithmetic intensity operations are typically referred to as streaming operations since they are typically bound by how quickly a processor can read and write data to its main memory. We propose a suite of new benchmark tests, which we call the Benchmark Streaming (BS) tests to allow us to focus on each of these distinct streaming operations. The suite consists of seven distinct vector operations: copy, AXPY, norm, inner product, fused CG update, gather, and scatter. We then describe an implementation of this suite of benchmarks using the OCCA portability library \cite{medina2014occa}. We show through computational experiments that we are able to obtain high memory bandwidth in each BS test on four different GPU accelerators: an Nvidia Titan V, an Nvidia V100, an AMD Radeon\TM VII, and an AMD Radeon Instinct\TM MI60. We also propose a simple performance model for the BS tests which we show can accurately model the observed bandwidths of each GPU considered.


It should be noted that memory bandwidth benchmarks for GPU accelerators is not a new proposal. Modern benchmarks such as Bablestream \cite{deakin2018evaluating}, Mixbench \cite{konstantinidis2017quantitative}, and SHOC \cite{danalis2010scalable} contain streaming kernel tests similar to some of the streaming tests we propose here. In particular, Babelstream considers `add', `triad', and `dot' tests which are similar to the AXPY and inner product tests we propose. Triad and a reduction test are also considered in SHOC. Each test we include in the BS suite, including ones similar to tests in other suites, is done for completeness, since each test we propose is relevant in the CG iterative method.



The remainder of this paper is organized as follows: we first give some brief details of the formulation of the CEED BPs in order to motivate the proposed gather and scatter streaming tests. We then detail our proposed Benchmark Streaming tests, and we propose a simple performance model of these streaming tests. Afterwards, we give some details of our open-source implementation of our proposed tests. Finally, we present performance data for each of the BS tests on four separate GPU accelerator devices, and conclude with summarizing remarks.

\section{CEED Benchmark Problems} \label{sec:CEED}

Here we give brief details regarding the formulation of the CEED BPs. While we do not consider the full benchmark problems in this paper, we introduce the basic finite element formulation in order to detail two specific streaming operations, namely the gather and scatter operations.

To begin, we consider an unstructured mesh of a domain $\mathcal{D} \subset \mathbb{R}^3$ into $K$ hexahedral elements $D^e$, where $e=1,\ldots,K$, such that
\begin{equation}
\label{eq:elSum}
\mathcal{D} = \bigcup_{e=1}^{K} D^e.
\end{equation}
We assume that each hexahedral element $D^e$ is an image of the reference bi-unit cube $\hat{D} = \{(r,s,t): -1 \leq r,s,t \leq1\}$ under a bijective map $\Gamma_e: \hat{D}\to D^e$. On each element $D^e$ we consider a set of nodes $\mathcal{N}_p^e = \{x_i^e,y_j^e,z_k^e\}_{i,j,k=0}^{p}$ defined as the image of the tensor product of $p+1$ Gauss-Lobatto-Legendre (GLL) nodes on reference cube, denoted $\hat{\mathcal{N}}_p = \{r_i,s_j,t_k\}_{i,j,k=0}^{p}$, under the map $\Gamma_e$.

On each element, we consider the space of polynomials of degree at most $p$ in each spatial dimension, and denote this space $\mathbb{Q}^p(D^e)$. As a polynomial basis of $\mathbb{Q}^p(D^e)$, we choose Lagrange interpolation polynomials, interpolating the nodal set $\mathcal{N}_p^e$. On the reference cube $\hat{D}$ we find that the basis of $\mathbb{Q}^p(\hat{D})$ can be written as a tensor product of the one-dimensional Lagrange interpolation polynomials $\{l_i\}_{i=0}^{i=p}$ interpolating the GLL nodes on the interval $[-1,1]$, that is, $\mathbb{Q}^p(\hat{D}) = \mathrm{span}(\{l_{ijk}(r,s,t)\}_{i,j,k=0}^{p})$ where
\begin{equation*}
    l_{ijk}(\mathbf{r}) = l_i(r)l_j(s)l_k(t),
\end{equation*}
for all $0\leq i,j,k \leq p$, and $\mathbf{r} = (r,s,t)$. Moreover, due to the construction of $\mathcal{N}_p^e$, the polynomial basis of $\mathbb{Q}^p(D^e)$ can be written as in terms of these reference basis polynomials. That is, given a $u^e \in \mathbb{Q}^p(D^e)$, we can write $u^e$ as
\begin{equation} \label{eq:basis}
    u^{e}\left(\mathbf{x}\right) = \sum_{i=0}^{p} \sum_{j=0}^p \sum_{k=0}^p u^e_{ijk} l_{ijk}\left(\Gamma_e^{-1}(\mathbf{x})\right),
\end{equation}
where $\mathbf{x} = (x,y,z)$. From to the definition of the Lagrange polynomials $l_i$ and the element mapping $\Gamma_e$, we find that the basis coefficients $u^e_{ijk}$ satisfy the interpolation property $u^e(x_i,y_j,z_k) = u^e_{ijk}$ for all $(x_i,y_j,z_k) \in \mathcal{N}_p^e$. We denote by $\mathbf{u}^e \in \mathbb{R}^{(p+1)^3}$ the vector of interpolation values $u^e_{ijk}$ on $D^e$.

On the full domain $\mathcal{D}$ we define the finite element space $V$ to be
\begin{equation*}
    V = \left\{v \in H^1(\mathcal{D}) \;\Big| \; v|_{D^e} \in \mathbb{Q}^p(D^e), e=1,\ldots,K\right\}.
\end{equation*}
From the interpolation property of the polynomial basis of $\mathbb{Q}^p(D^e)$ on each element $D^e$, we see that a function $v \in V$ is completely determined by its nodal values at each of the unique interpolation nodes in the mesh of $\mathcal{D}$. Let us denote this set of all the unique interpolation nodes in $\mathcal{D}$ by $\mathcal{N}_G$ and define $N_G = |\mathcal{N}_G|$ to be the total number of interpolation nodes in the mesh. For a given $v \in V$ we can therefore represent $v$ as the vector $\mathbf{v}_G \in \mathbb{R}^{N_G}$ of the values of $v$ at each of the interpolation nodes in $\mathcal{N}_G$.

It is also useful to represent a function $v \in V$ as a larger vector of length $N_L = K(p+1)^3$ whose entries are ordered by the elements $D^e$ in the mesh of $\mathcal{D}$. That is, we represent $v$ as a vector $\mathbf{v}_L \in \mathbb{R}^{N_L}$, such that $\mathbf{v}_L = (\mathbf{v}^1, \ldots, \mathbf{v}^K)^T$. Each entry of $\mathbf{v}_L$ therefore corresponds to distinct entry in $\mathbf{v}_G$ and the two vectors can be connected via a \emph{scatter} operator, $Z$:
\begin{equation} \label{eq:scatter_def}
    \mathbf{v}_L = Z \mathbf{v}_G.
\end{equation}
The matrix $Z$ is therefore an $N_L \times N_G$ Boolean matrix with each row containing precisely one nonzero. The transpose of the scatter operator, $Z^T$, often referred to as the \emph{gather} operator. Both operators are described in more detail in \cite{canuto2012spectral}, \cite{deville2002high}, and \cite{henderson1995unstructured}.

The specific finite element operators considered in each of the CEED benchmark problems consist of some variant of the problem of finding $u \in V$ such that
\begin{equation} \label{eq:helm_var}
    \int_{\mathcal{D}} \nabla v^T  \nabla u  + \lambda v u\; d\mathbf{x} = \int_{\mathcal{D}} v f \; d\mathbf{x}
\end{equation}
for all $v \in V$, given a function $f \in V$. After discretizing via the finite element method, the details of which we omit here, the global problem \eqref{eq:helm_var} can be written as the following system for a vector of global degrees of freedom $\mathbf{u}_G$,
\begin{equation}\label{eq:global_system}
    S_G \mathbf{u}_G + \lambda M_G \mathbf{u}_G = \mathbf{b}_G,
\end{equation}
where the action of the global stiffness matrix $S_G$ and mass matrix $M_G$ are formed by the composition of scattering, local stiffness/mass action, and gathering operations. That is, the following forms hold
\begin{align}
S_G &= Z^T S_L Z, \label{eq:assembled_stiff}\\
M_G &= Z^T M_L Z, \label{eq:assembled_mass}
\end{align}
where $S_L$ and $M_L$ are block diagonal matrices whose blocks contain non-zeros only on the degrees of freedom local to each element.

\begin{algorithm*}[t]
  \caption{Conjugate Gradient Method}
  \label{alg:cg}
\begin{boxedminipage}{\textwidth}
    \begin{algorithmic}[1]
    \STATE {\bf Input:} (1) Initial guess $\mathbf{x}$, (2) Right-hand-side vector $\mathbf{b}$, (3) Linear operator $A$, (4) Tolerance $\epsilon$.
    \STATE {{\bf Output:} (1) Solution $\mathbf{x}$, (2) Iteration count $j$}
    \STATE {Set $j=0$}
    \STATE {Set $\mathbf{r}_0 = \mathbf{b}-A\mathbf{x}$}
    \STATE {Set $\mathbf{p} = \mathbf{r}$}
    \WHILE {($\mathbf{r}_j\cdot\mathbf{r}_j > \epsilon$)}
        \STATE {$\alpha = \frac{\mathbf{r}_j\cdot\mathbf{r}_j}{\mathbf{p}\cdot A\mathbf{p}}$}
        \STATE {$\mathbf{x} = \mathbf{x} + \alpha\mathbf{p}$}
        \STATE {$\mathbf{r}_{j+1} = \mathbf{r}_j - \alpha A\mathbf{p}$}
        \STATE {$\beta = \frac{\mathbf{r}_{j+1}\cdot\mathbf{r}_{j+1}}{\mathbf{r}_j\cdot\mathbf{r}_j}$}
        \STATE {$\mathbf{p} = \mathbf{r}_{j+1} + \beta \mathbf{p}$}
        \STATE {$j=j+1$}
    \ENDWHILE
  \end{algorithmic}
\end{boxedminipage}
\end{algorithm*}

It is common in some high-order element software to consider the system \eqref{eq:global_system} instead in a scattered form by applying the scatter operator $Z$ to both sides of \eqref{eq:global_system}, and solving for the scattered degrees of freedom, recalling the definition of the scattered vector of degrees of freedom in \eqref{eq:scatter_def}. However, storing the full finite element problem in the scattered degrees of freedom $\mathbf{u}_L$ will require significantly more data movement compared to the considerably smaller vector $\mathbf{u}_G$. Since we do not consider the solution of the full finite element problem \eqref{eq:global_system} in this paper, we will instead investigate the performance of both the gather and scatter operators to investigate the cost of moving between gathered and scattered degree of freedom storage.

\section{Benchmark Streaming Tests} \label{sec:BS}

Each of the CEED benchmark problems involves performing a Conjugate Gradient iterative method. We list the pseudo-code for this iterative method in Algorithm \ref{alg:cg}. The method consists of the following simple vector operations:
\begin{itemize}
    \item Vector addition: $\mathbf{y} = \alpha \mathbf{x} + \beta \mathbf{y}$
    \item Dot product: $\alpha = \mathbf{x} \cdot \mathbf{y}$
    \item Norm: $\alpha = \mathbf{x} \cdot \mathbf{x}$
    \item Matrix-vector Product: $\mathbf{y} = A\mathbf{x}$
\end{itemize}
The first three operations perform only a small number of floating point operations per vector value loaded, and therefore their performance is limited by the rate at which vector data can be streamed to and from main memory. Likewise, though the element local blocks of the finite element operators $S_L$ and $M_L$ are dense, the transition between global and local degrees of freedom via the scatter operator $Z$ in \eqref{eq:assembled_stiff}-\eqref{eq:assembled_mass} is of similarly low arithmetic intensity. We therefore refer to these operations as \emph{streaming} operations.

\begin{table}[t]
\begin{center}
    \begin{tabular}{|c|c|c|} \hline
    Test & Name & Description \\ \hline
    BS1 & Vector Copy & $\mathbf{y} = \mathbf{x}$  \\
    BS2 & Vector AXPY & $\mathbf{y} = \alpha\mathbf{x} + \beta\mathbf{y}$ \\
    BS3 & Vector Norm & $\alpha = \mathbf{x} \cdot \mathbf{x}$\\
    BS4 & Vector Inner Product & $\alpha = \mathbf{x} \cdot \mathbf{y}$\\
    BS5 & Fused CG Update & $\left\{ \begin{aligned} \mathbf{x} &= \mathbf{x} + \alpha\mathbf{p}, \\ \mathbf{r} &= \mathbf{r} - \alpha A\mathbf{p}, \\ \beta &= \mathbf{r}\cdot\mathbf{r}. \end{aligned} \right.$ \\
    BS6 & Gather  & $\mathbf{x}_G = Z^T\mathbf{x}_L$\\
    BS7 & Scatter & $\mathbf{x}_L = Z\mathbf{x}_G$\\\hline
    \end{tabular}
    \end{center}
    \caption{Summary of Benchmark Streaming problems.}
    \label{bs.tab}
\end{table}

To help investigate the performance and portability of the relevant streaming operations in the CEED benchmarks, we propose a suite of Benchmark Streaming (BS) problems, listed in Table \ref{bs.tab}. These BS problems consists of individual tests for the following vector operations: copy, AXPY, norm, inner product, fused CG update, gather, and scatter. Each of the BS problems is motivated by vector operations performed in the CG iterative method. Note that the fused CG update originates from noticing that steps 8-10 of the CG method in Algorithm \ref{alg:cg} can be fused into a single operation, which helps save data movement from an additional read/write of $\mathbf{r}$.

The relevant figure of merit in each of the BS problems should be data throughput, measured in GB/s. To eliminate sources of variance between different vendors' implementations of device timers in heterogeneous compute models, we measure all time with respect to the host process with a timer such as {\verb MPI_Wtime }.

The first 5 BS problems are simple vector operations and thus are more flexible in that they can be performed in absence of any knowledge of an underlying finite mesh. BS6 and BS7, however, require knowledge of the finite element mesh via the scatter operator $Z$. For these problems, we specify the same configuration as detailed for the CEED BPs above. We note that although the gather/scatter operators in BS6 and BS7 could have a non-trivial parallel implementation with MPI, we restrict our attention to solely to local computations on a single MPI rank in this paper.

\section{Performance Model} \label{sec:model}

\begin{lstlisting}[float=*,language=C++,escapeinside=??, caption={BS1: entry-wise copy of an array.}, label={BS1Kernel.okl}]
// BS1: Copy an array
?@?kernel void BS1(const int N,     // number of entries in array
                 ?@?restrict const double *a, // input:  device array
                 ?@?restrict double *b){      // output: device array

  // parallel loop blocked over thread-blocks of size p_BLOCKSIZE
  for(int n=0;n<N;++n;?@?tile(p_BLOCKSIZE,?@?outer,?@?inner)){
    b[n] = a[n];
  }
}
\end{lstlisting}

\begin{lstlisting}[float=*,language=C++,escapeinside=??, caption={BS2: Scaled in-place addition of two arrays.}, label={BS2Kernel.okl}]
// BS2: AXPY
?@?kernel void BS2(const int N,  // number of entries in array
                 const double alpha,          // input:  scalar
                 ?@?restrict const double *x,  // input:  device array
                 const double beta,           // input:  scalar
                 ?@?restrict double *y){       // output: device array

  // parallel loop blocked over thread-blocks of size p_BLOCKSIZE
  for(int n=0;n<N;++n;?@?tile(p_BLOCKSIZE,?@?outer,?@?inner)){
    y[n] = alpha*x[n]+beta*y[n];
  }
}
\end{lstlisting}

\begin{lstlisting}[float=*,language=C++,escapeinside=??, caption={BS3: Simple partial reduction kernel for computing vector norm.}, label={BS3Kernel.okl}]
// BS3_1: 1st reduction kernel for computing vector norm
?@?kernel void BS3_1(const int Nblocks,  // num entries in partially-reduced array
                   const int N,        // number of entries in array
                   ?@?restrict const  double *a, // input:  device array
                   ?@?restrict double *norm_tmp){// output: partially-reduced norm

  // parallel loop over of Nblocks work-groups
  for(int b=0;b<Nblocks;++b;?@?outer(0)){

    ?@?shared volatile double s_norm[p_BLOCKSIZE];

    for(int t=0;t<p_BLOCKSIZE;++t;?@?inner(0)){
      s_norm[t] = 0.0;
      for (int id=t+b*p_BLOCKSIZE; id<N; id+=p_BLOCKSIZE*Nblocks) {
        s_norm[t] += a[id]*a[id];
      }
    }

    //block reduction in shared memory
    //assumes p_BLOCKSIZE is a power of 2, and >=2
    for(int k=p_BLOCKSIZE/2;k>1;k>>=1){
      for(int t=0;t<p_BLOCKSIZE;++t;?@?inner(0)) {
        if(t<k) s_norm[t] += s_norm[t+k];
      }
    }

    for(int t=0;t<p_BLOCKSIZE;++t;?@?inner(0))
      if(t<1) norm[b] = s_norm[0] + s_norm[1];
  }
}
\end{lstlisting}

\begin{lstlisting}[float=*,language=C++,escapeinside=??, caption={BS6: Gather operator.}, label={BS6Kernel.okl}]
// BS6: Gather
?@?kernel void BS6(const int Nblocks,   // number of blocks in grid
                 ?@?restrict const int *blockStarts, // offset array blocking rows
                 ?@?restrict const int *rowStarts,// row starts in csr data struct
                 ?@?restrict const int *colIds,   // column ids in csr data struct
                 ?@?restrict const double *q,     // input:  device array
                 ?@?restrict double *gatherq) {   // output: gathered device array

  for(int b=0;b<Nblocks;++b;?@?outer(0)){
    ?@?exclusive int blockStart, blockEnd;
    ?@?shared double temp[p_NODES_PER_BLOCK];

    //read all data needed to gather a block of rows
    for(int n=0;n<p_BLOCKSIZE;++n;?@?inner(0)){
      blockStart = blockStarts[b];
      blockEnd   = blockStarts[b+1];

      const int row = rowStarts[blockStart]+n;
      //checking for the last block lets the compiler unroll the load
      if (b!=Nblocks-1)
        for (int i=0;i<p_NODES_PER_BLOCK;i+=p_BLOCKSIZE)
          temp[i+n] = q[colIds[i+row]];
      else
        for (int i=0;row+i<rowStarts[blockEnd];i+=p_BLOCKSIZE)
          temp[i+n] = q[colIds[i+row]];
    }

    // gather values inside shared memory, and write out
    for(int n=0;n<p_BLOCKSIZE;++n;?@?inner(0)){
      for (int row = n+blockStart; row<blockEnd; row+=p_BLOCKSIZE) {
        const int colStart = rowStarts[row]  - rowStarts[blockStart];
        const int colEnd   = rowStarts[row+1]- rowStarts[blockStart];
        double gq = 0.0;
        for (int i=colStart;i<colEnd;i++)
          gq += temp[i];

        gatherq[row] = gq;
      }
    }
  }
}
\end{lstlisting}

\begin{lstlisting}[float=*,language=C++,escapeinside=??, caption={BS7: Scatter operator.}, label={BS7Kernel.okl}]
// BS7: Scatter
?@?kernel void BS7(const int N,  // number of entries in scattered array
                 ?@?restrict const int *scatterIds, // ids of scattered values
                 ?@?restrict const double *gatherq, // input: gathered array
                 ?@?restrict double *q) {           // output: scattered array

  for(int n=0;n<N;++n;?@?tile(p_BLOCKSIZE, ?@?outer(0), ?@?inner(0))){
    const int id = scatterIds[n];
    if(id>=0)
      q[n] = gatherq[id];
  }
}
\end{lstlisting}

The goal of the streaming tests proposed in this paper is not simply to obtain a measure of a processor's maximum bandwidth in these simple vector operations, but to provide valuable insight into modeling such operations when estimating factors such as weak- and strong-scaling efficiency. To this end, we propose here a simple performance model of these streaming kernels, within which are two useful performance parameters.

In most modern fork \& join programming paradigms, for example offloading to a GPU accelerator, it is common for there to be some measure of `launch latency'. That is, there is some non-zero time required simply to offload work to an accelerator, or fork and join a group of threads. Let us denote the time it takes to stream $B$ bytes of data as $T(B)$. A simple but effective model for $T(B)$ consists of this minimum time it will take to launch a streaming kernel, $T_0$, plus the time it will take to stream $B$ bytes. Assuming that the time it takes to stream $B$ bytes is simply a linear function of the amount of data $B$, a simple Amdahl-type model for $T_B$ can be written as
\begin{equation}\label{eq:time_model}
T(B) = T_0 + \frac{B}{W_{\max}}.
\end{equation}
The parameter $W_{\max}$ is therefore the asymptotic maximum bandwidth that can be attained by the kernel for streaming data to and from the device's main memory. The effective bandwidth achieved by the device when streaming $B$ bytes of data would therefore be modeled as
\begin{align}
    W_{\mathrm{eff}}(B) &= \frac{B}{T(B)}, \nonumber \\
                     &= \frac{B}{T_0 + \frac{B}{W_{\max}} }. \label{eq:perf_model}
\end{align}
Fitting this model to the measured bandwidth in each of the streaming tests therefore yields several useful estimates. First, $T_0$ gives us an estimate of the fixed launch cost. Second, $W_{\max}$ gives us an estimate for the asymptotic peak bandwidth which the device can effectively achieve. Together, these parameters used in \eqref{eq:perf_model} can give effective timing/bandwidth estimates for different sized buffers being streamed. For example, it can be useful to know precisely how much data must be moved by a kernel for the device to be operating at 80\% of its theoretical maximum throughput. This 80\% efficiency size in bytes, which we denote by $B_{0.8}$, has been termed the Fischer-Brown point. This value can be predicted by our performance model as follows
\begin{align}
    W_{\mathrm{eff}}(B_{0.8}) &= 0.8 W_{\max}, \nonumber \\
    \frac{B_{0.8}}{T_0 + \frac{B_{0.8}}{W_{\max}} } &= 0.8 W_{\max}, \nonumber \\
    \Rightarrow B_{0.8} &= 4 T_0 W_{\max}. \label{eq:Fischer-Brown}
\end{align}
Since the product $ T_0 W_{\max}$ is a measure of how much data the device could have been theoretically moving in the time taken to offload and sync, this result matches intuition in saying that to reach 80\% efficiency we should move four times this amount of data in the kernel.

\section{Implementation} \label{sec:implementation}

We implement each of the benchmark streaming tests described above in C++, using a stripped-down version of the open-source finite-element library libParanumal \cite{libparanumal2020} as a base to perform mesh generation and handle data structures. The source code for each of our BS tests is available freely in the streamParanumal GitHub repository \cite{streamparanumal2020}.

The libParanumal core of our benchmark implementation uses the Open Concurrent Compute Abstraction (OCCA) library \cite{medina2014occa} to abstract between different parallel languages such as OpenMP, OpenCL\TM, CUDA, and HIP. OCCA allows us to implement the parallel kernel code for each of the BS tests in a (slightly decorated) C++ language, called OKL. At runtime, the user can specify which parallel programming model to target, after which OCCA translates the OKL source code into the desired target language and Just-In-Time (JIT) compiles kernels for the user's target hardware architecture.

In the OKL language, parallel loops and variables in special memory spaces are described with simple attributes. For example, iterations of nested parallel for loops in the kernel are annotated with \texttt{@outer} and \texttt{@inner} to describe how they are to be mapped to a grid of work-item and work-groups in OpenCL\TM or threads and thread-blocks in CUDA and HIP. All iterations that are annotated with \texttt{@outer} or \texttt{@inner} are assumed to be free of loop carried dependencies.

The compute kernels for BS1 and BS2 are relatively short, given the simple nature of both benchmarks. We list the OKL source for BS1 and BS2 in Listings 1 and 2, respectively. These kernels can be written compactly using OCCA's \texttt{@tile} attribute, which instructs the loop to be parallelized over a grid of blocks of size {\verb p_BLOCKSIZE }, which we typically take to be 256 or 512. The BS3-5 benchmarks, on the other hand, each require performing some sort of reduction. These benchmarks are implemented using two compute kernels each, the first of which performs a partial reduction of the needed array, while the second kernel completes the reduction to a single value. We list a particularly simple example of a partial reduction performed for computing the vector norm for BS3 in Listing \ref{BS3Kernel.okl}. This example OKL kernel reduces the norm of the entire vector to an array of \texttt{Nblocks} entries, which we typically take to be some relatively small value such as 512. The second kernel which reduces this partially reduced array to a single value would be written analogously.

The implementation of a reduction in Listing \ref{BS3Kernel.okl} performs quite well for each GPU architecture we consider in our computational tests below. However, it is not necessarily the only implementation of a reduction used in our BS tests. Often, small improvements in the asymptotic maximum achieved bandwidth ($W_{\max}$ in our performance model) can be obtained by implementing reductions by completing the last warp/wavefront-wide reduction in register memory using intrinsic { \verb shfl } instructions (or { \verb shfl_sync } in the case of CUDA 9+). These special reduction implementations can be specified based on our target device at JIT compilation time. For brevity, we do not list each of these specialized reductions in this paper, nor the OKL kernel source code for the BS4 and BS5 tests, but they can be found in the source code for each of the BS tests in \cite{streamparanumal2020}.

To implement the BS6 and BS7 tests we require several data structures to be constructed during mesh setup. In particular, we require some sort of sparse matrix representation of the scatter and gather operators, $Z$ and $Z^T$, respectively, described above. In the libParanumal mesh setup code, we simplify how these sparse matrices are stored by noting that each entry of $Z$ is 1, and that $Z$ has one non-zero entry per row. This allows us to implement the gather and scatter operators for BS6 and BS7 with very specialized sparse matrix multiplication compute kernels.

We list the OKL kernel source for BS6 in Listing \ref{BS6Kernel.okl}. This kernel's implementation is inspired by the study of optimized sparse CSR matrix-vector multiplication on GPU accelerators in \cite{greathouse2014efficient}. In particular, we assume that, in addition to representing the non-zero entries of the gather operator $Z^T$ in sparse CSR format (omitting the array of matrix values, since they are all ones), we have constructed an array of offsets into the array of row starts, which we call the \texttt{blockStarts} arrays. We populate this array such that the number of rows to be processed by each thread-block is approximately balanced. Specifically, rows are blocked in such a way that each thread-block reads approximately { \verb p_NODES_PER_BLOCK } entries in the sparse CSR matrix into shared memory, where the actual matrix action is computed. We typically select {\verb p_BLOCKSIZE } to be 256 and { \verb p_NODES_PER_BLOCK } to be some small multiple of {\verb p_BLOCKSIZE }, such as 512. This allows each thread-block to read and write approximately the same amount of data, in an efficient manner. Since the matrix $Z^T$ has at most one non-zero per column, there is relatively little opportunity for effective cache utilization when reading in the sparse entries of \texttt{q}, but spatial locality of the local DOF storage in \texttt{q} may help somewhat.

Finally, we list the OKL kernel source for BS7 in Listing \ref{BS7Kernel.okl}. In comparison to the gather operator kernel, this kernel is quite simple. This is because in this kernel we can leverage the fact that each row of the scatter operator $Z$ has one non-zero entry in order to compute a mapping array \texttt{scatterIds} which maps each of the local storage DOFs to its index in the global DOF array. We reserve negative values in the \texttt{scatterIds} array for the case where degrees of freedom are masked out of the scatter operator action, say through enforcement of boundary conditions.


\section{Computational study} \label{sec:compStudy}

\begin{table*}[th!]
    \centering
    \begin{tabular}{|l||c|c|cccc|c|} \cline{4-8}
     \multicolumn{3}{c|}{}& \multicolumn{4}{c|}{Aggregate memory (MB)} & \multicolumn{1}{c|}{Device memory}\\ \hline
     \multicolumn{1}{|c||}{Model} & \multicolumn{1}{c|}{TDP}    & \multicolumn{1}{c|}{CUs}  & \multicolumn{1}{c}{REG} & \multicolumn{1}{c}{LDS} & \multicolumn{1}{c}{L1}  & \multicolumn{1}{c|}{L2} & \multicolumn{1}{c|}{HBM2} \\ \hline
     AMD Radeon\TM VII & 250 W & 60 & 15 & 3.75 & 0.9 & 4   & 16 GB \\
     AMD MI60       & 225 W & 64 & 16 & 4    & 1   & 4   & 32 GB \\ \hline
     Nvidia TitanV      & 250 W & 80 & 20 & 7.5  & 2.5 & 4.5 & 12 GB \\
     Nvidia V100        & 300 W & 80 & 20 & 7.5  & 2.5 & 6   & 32 GB \\  \hline
    \end{tabular}
    \caption{Specifications for four contemporary graphics processing units. Here: CU stands for the number of compute units on AMD GPUs or the number of streaming multiprocessors on Nvidia GPUs, REG is the aggregate size of the register files, LDS is the aggregated amount of local shared cache, L1 is the aggregate amount of lowest level cache, L2 is the total amount of second level cache, and HBM2 is the amount of high-bandwidth memory.}
    \label{gpuSpecs.tab}
\end{table*}

In this section, we present results from our implementation of the BS benchmarks on four current generation GPU accelerators from two manufacturers. We demonstrate that each of our tests is able to obtain reasonably high throughput of device global HBM memory, and that our performance model detailed above can fit the observed performance on each of the considered devices very well.

\subsection{Testing Platforms}
The four contemporary GPU accelerator devices considered in this paper are the Titan V and V100 GPUs manufactured by Nvidia, and the Radeon\TM VII and Radeon Instinct\TM MI60 GPUs manufactured by AMD. In Table \ref{gpuSpecs.tab} we show some physical specifications for each of the considered GPUs including their thermal design power (TDP), number of stream multiprocessors/compute units (CUs), aggregate register space (REG), aggregate shared memory/local data storage (LDS), aggregate default lowest level cache (L1), aggregate second level cache size (L2), and amount of second generation high bandwidth memory (HBM2).

The most important performance metric of each of the GPUs considered for their overall performance in the BS benchmarks will be the delivered bandwidth to global HBM memory. While the physical specification of the Nvidia V100 and Titan V GPUs differ only in their TDP and HBM2 amounts, the former has four 8 GB HBM2 memory stacks and the latter only has three 4 GB HBM2 memory stacks. This leads to a difference in theoretical peak HBM bandwidth between the two devices, with the V100 having a theoretical peak HBM bandwidth of 900 GB/s and the Titan V having a theoretical peak HBM bandwidth of 653 GB/s. Since each memory stack has associated L2 cache, this also leads to the difference in L2 cache size.

By contrast, the AMD Radeon\TM VII and Radeon Instinct\TM MI60 each have four HBM2 memory stacks, the former using 4 GB stacks and the latter 8 GB stacks. Both GPUs therefore have a theoretical peak HBM bandwidth of 1000 GB/s. Despite this advantage in theoretical bandwidth over the Nvidia GPUs, it is important to note the AMD GPUs also have comparatively fewer compute units, lower aggregate register count, smaller aggregate LDS space, smaller aggregate L1 cache, and a smaller L2 cache. These variations between the different GPU architectures considered here adds complexity when attempting to directly compare the actual measured performance of a compute task.

The Nvidia Titan V test system used in this paper contained two {Intel Xeon Gold 6128 CPU @ 3.40GHz} CPUs, running {Ubuntu 18.04} and {CUDA 9.1}. The Nvidia V100 system was run via an Amazon Web Services (AWS) instance on a system containing an {Intel Xeon CPU E5-2686v4 @ 2.30GHz} CPU, running {Ubuntu 16.04} and {CUDA 10.2}. The AMD Radeon\TM VII system used here contained an {Intel Xeon CPU E5-1650 v3 @ 3.50GHz} CPU, running {Ubuntu 18.04} and {ROCm 3.5}. The AMD Radeon Instinct\TM MI60 system was accessed as a node of the Lawrence Livermore National Laboratory (LLNL) Corona cluster and contained an {AMD EYPC\TM 7401} CPU @ 2.00GHz, running {TOSS 3} and {ROCm 3.5}.

\subsection{Experiments}

\begin{figure*}[th!]
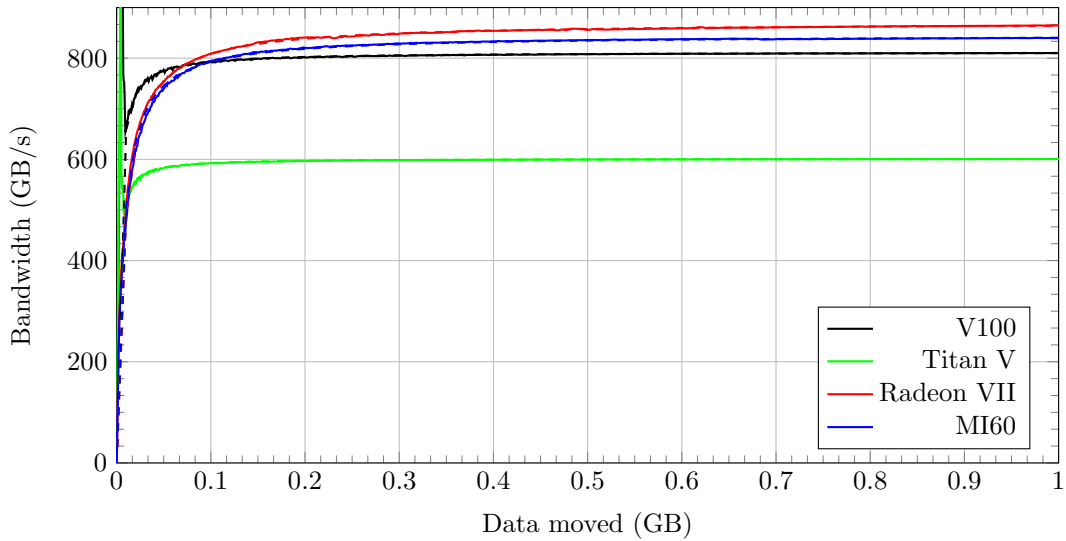

\begin{center}

\end{center}
\caption{Results for BS1. Colored solid lines show the measured bandwidth for the Nvidia Titan V, Nvidia V100, AMD Radeon\TM VII, and AMD Radeon Instinct\TM MI60. Colored dashed lines show the performance models for each of the GPU accelerators.}
\label{BS1.fig}
\end{figure*}

\begin{figure*}[th!]
\begin{center}
%
\end{center}
\caption{Results for BS2. Colored solid lines show the measured bandwidth for the Nvidia Titan V, Nvidia V100, AMD Radeon\TM VII, and AMD Radeon Instinct\TM MI60. Colored dashed lines show the performance models for each of the GPU accelerators.}
\label{BS2.fig}
\end{figure*}

\begin{figure*}[htbp!]
\begin{center}
%
\end{center}
\caption{Results for BS3. Colored solid lines show the measured bandwidth for the Nvidia Titan V, Nvidia V100, AMD Radeon\TM VII, and AMD Radeon Instinct\TM MI60. Colored dashed lines show the performance models for each of the GPU accelerators.}
\label{BS3.fig}
\end{figure*}

\begin{figure*}[htbp!]
\begin{center}
%
\end{center}
\caption{Results for BS4. Colored solid lines show the measured bandwidth for the Nvidia Titan V, Nvidia V100, AMD Radeon\TM VII, and AMD Radeon Instinct\TM MI60. Colored dashed lines show the performance models for each of the GPU accelerators.}
\label{BS4.fig}
\end{figure*}

\begin{figure*}[htbp!]
\begin{center}
%
\end{center}
\caption{Results for BS5. Colored solid lines show the measured bandwidth for the Nvidia Titan V, Nvidia V100, AMD Radeon\TM VII, and AMD Radeon Instinct\TM MI60. Colored dashed lines show the performance models for each of the GPU accelerators.}
\label{BS5.fig}
\end{figure*}

\begin{figure*}[th!]
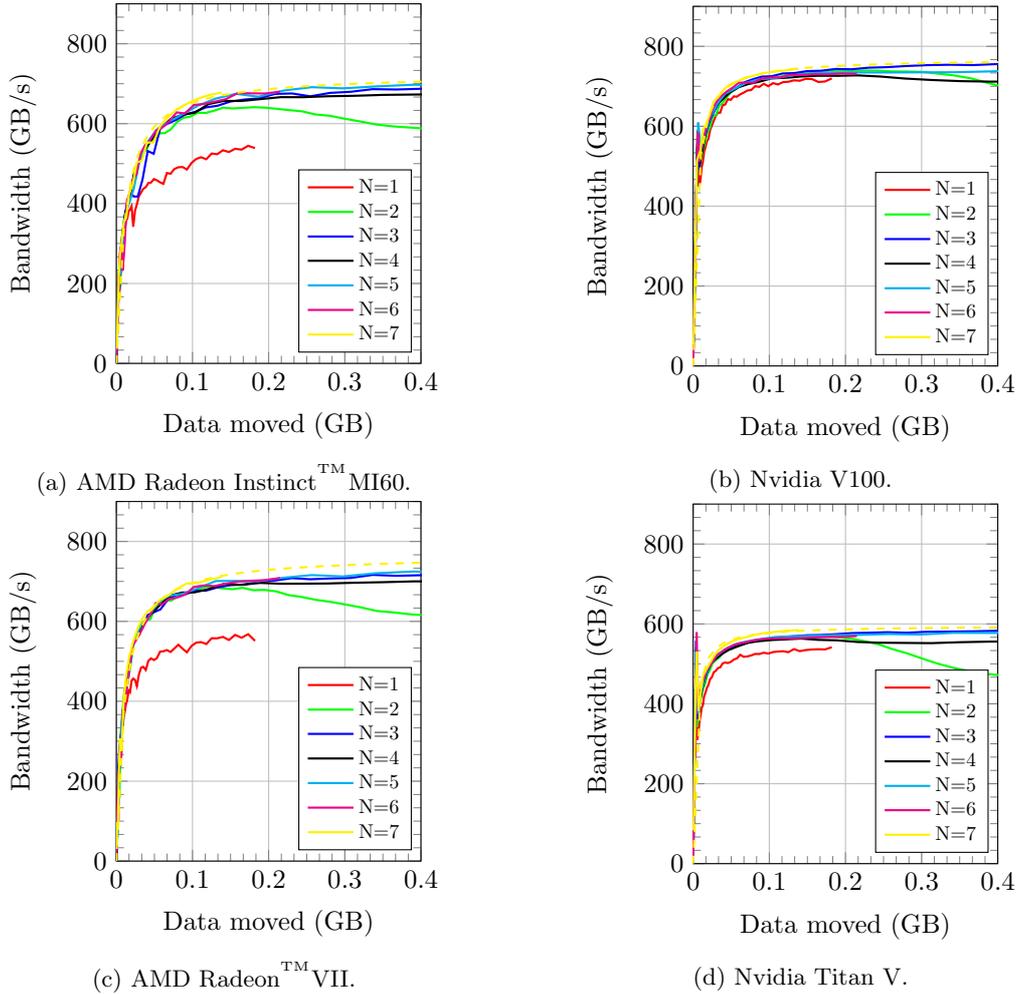

\begin{subfigure}{.49\textwidth}
  \centering
%
\caption{AMD Radeon Instinct\TM MI60.}
\label{BS6_MI60.fig}
\end{subfigure}
\begin{subfigure}{.49\textwidth}
  \centering
%
\caption{Nvidia V100.}
\label{BS6_V100.fig}
\end{subfigure}

\begin{subfigure}{.49\textwidth}
  \centering
%
\caption{AMD Radeon\TM VII.}
\label{BS6_R7.fig}
\end{subfigure}
\begin{subfigure}{.49\textwidth}
  \centering
%
\caption{Nvidia Titan V.}
\label{BS6_TITANV.fig}
\end{subfigure}

\caption{Results for BS6. Colored solid lines in each sub-figure show the measured bandwidth for polynomial degrees $N=1,\ldots,7$.}
\label{BS6.fig}

\end{figure*}

\begin{figure*}[th!]
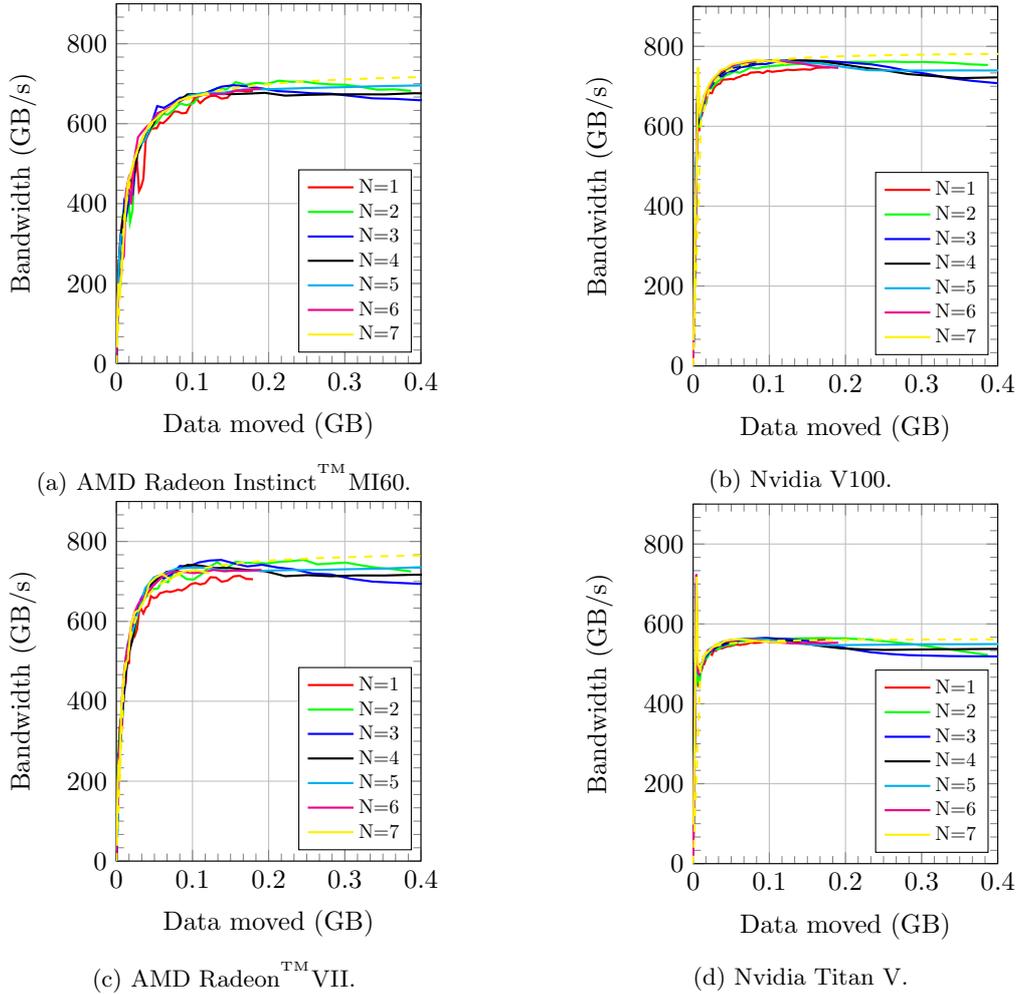

\begin{subfigure}{.49\textwidth}
  \centering
%
\caption{AMD Radeon Instinct\TM MI60.}
\label{BS7_MI60.fig}
\end{subfigure}
\begin{subfigure}{.49\textwidth}
  \centering
%
\caption{Nvidia V100.}
\label{BS7_V100.fig}
\end{subfigure}

\begin{subfigure}{.49\textwidth}
  \centering
%
\caption{AMD Radeon\TM VII.}
\label{BS7_R7.fig}
\end{subfigure}
\begin{subfigure}{.49\textwidth}
  \centering
%
\caption{Nvidia Titan V.}
\label{BS7_TITANV.fig}
\end{subfigure}

\caption{Results for BS7. Colored solid lines in each sub-figure show the measured bandwidth for polynomial degrees $N=1,\ldots,7$. The single dashed line in each plot shows the performance model fit for $N=7$.}
\label{BS7.fig}

\end{figure*}

We begin with the BS1-5 test results shown in Figures \ref{BS1.fig}-\ref{BS5.fig}. In each figure, we plot the effective bandwidth achieved by each GPU device on the $y$ axis in GB/s with total data moved by the test in GB on the $x$ axis. To collect the data for each BS1-5 test, we record the achieved bandwidth by running the relevant compute kernel 20 times and dividing the total amount of data moved by the total time to run the 20 kernel invocations. This helps to reduce small variations in each individual kernel run time. For each plot, we run the relevant benchmark for over 400 different array sizes using each GPU device and plot the achieved bandwidth in GB/s, for each device. The data for each GPU appears as a generally smooth curve in each of the plots.

In each of the BS1-5 plots in Figures \ref{BS1.fig}-\ref{BS5.fig}, though it is often difficult to perceive, we also plot the model fit in \eqref{eq:perf_model} for the measured bandwidth of each GPU device as a colored dashed line. The model fit is difficult to perceive because it fits the experimental data extremely well and is often laying directly on top of the curve made by the experimental measurements. To fit the performance model \eqref{eq:perf_model} to the experimental data, we do a simple least squares linear fit of the elapsed time of each trial, effectively finding the $T_0$ and $W_{\max}$ parameters through the linear model \eqref{eq:time_model}. We summarize each of the model parameters for each of the BS1-7 tests and for each device in Table \ref{BSparams.tab} at the end of this section.

Some general trends can be observed across Figures \ref{BS1.fig}-\ref{BS5.fig}. First, both the Nvidia V100 and Titan V are observed to have a spike in measured bandwidths at small transfer sizes. The peaks of these spikes precisely correspond to the size of the L2 cache on both devices, and manifest because of the consecutive runs of the kernel in each tests. It is interesting to note that this cache effect is most pronounced in BS1 and BS2, and is significantly less pronounced in BS3, BS4, and BS5, indicating that there may be some sort of cache invalidation occurring by BS3-5 launching two distinct kernels in each iteration. By contrast, despite their comparably sized L2 cache to the Nvidia Titan V, the AMD Radeon\TM VII and Radeon Instinct\TM MI60 appear to show no analogous cache effects. This suggests the L2 cache is invalidated between each kernel launch on the AMD GPUs.

In general, each of the Nvidia V100, AMD Radeon\TM VII, and AMD Radeon Instinct\TM MI60 achieve an asymptotic maximum bandwidth near 800 GB/s, while the Nvidia Titan V achieves slightly more than 600 GB/s due to its smaller number of HBM2 memory stacks. The curious exception to this is in BS4 where the Nvidia V100 achieves a significantly higher asymptotic maximum bandwidth than it does in either BS3 or BS5, which are similar kernels to BS4. The AMD Radeon\TM VII also performs slightly worse in BS4 and BS5, achieving only near 750 GB/s asymptotic maximum bandwidth.

While the AMD Radeon\TM VII and Radeon Instinct\TM MI60 achieve an asymptotic maximum bandwidth near or higher than the Nvidia V100 in each of our BS tests, it takes significantly more data being moved by each GPU to achieve near their maximum bandwidths. This is captured in our performance model \eqref{eq:perf_model} by a larger fixed launch cost $T_0$. In the summarized model parameters given in Table \ref{BSparams.tab}, we see that the fit for $T_0$ in BS1 and BS2 for the AMD Radeon\TM VII and Radeon Instinct\TM MI60 is roughly 9 $\mu$s, and nearly triple that of the Nvidia V100 model's $T_0$ values. Recalling the definition of the Fischer-Brown point $B_{0.8}$ in \eqref{eq:Fischer-Brown} (i.e., the amount of data necessary to transfer to achieve 80\% of maximum bandwidth), we see that the larger $T_0$ parameters directly impacts the size of $B_{0.8}$, requiring more data to be moved to achieve 80\% efficiency. For example, for the norm calculation in BS3, the Fischer-Brown point for the Nvidia V100 is $B_{0.8} = 24$ MB, which for BS3 implies the vector being streamed must be at least 3.16M doubles. By comparison, the Fischer-Brown point for the AMD Radeon Instinct\TM MI60 is $B_{0.8} = 55.9$ MB, which for BS3 implies the vector being streamed is at least 7.33M doubles.

Finally, we show the results of BS6 and BS7 in Figures \ref{BS6.fig} and \ref{BS7.fig}, respectively. In both tests, we show results for each of the four GPUs considered for hexahedral finite meshes of order $N=1,\ldots, 7$. Each figure shows the bandwidth measured in each test for a chosen polynomial order $N$ over several mesh sizes. The meshes are chosen to be structured $K\times K \times K$ meshes of hexahedral elements, for various values of $K$, beginning at $K=2$. Though the meshes are chosen to be structured, we make no use of that knowledge in the mesh data structures, nor in the scatter and gather operators, $Z$ and $Z^T$, respectively. For each mesh size and polynomial order, we plot the measured bandwidth for the test as a function of the total amount of data moved in the kernel on the $x$ axis. We also plot in each figure the fit performance model for the $N=7$ data. We see in each figure that the performance of the gather and scatter kernels follow roughly the same performance trend for each considered polynomial order. The exception to this is the case of $N=1$ and $N=2$ in BS6, where the performance of the AMD Radeon\TM VII and Radeon Instinct\TM MI60 along with the Nvidia Titan V appears to be suboptimal. The Nvidia V100, on the other hand, performs consistently for all considered polynomial orders, potentially due to its large cache sizes compared with the other GPUs.

\begin{table*}[t]
    \centering
    \begin{tabular}{|l||c|c||c|c||c|c||c|c|}
     \hline
     \multicolumn{1}{|c||}{} & \multicolumn{2}{c||}{AMD Radeon\TM VII} & \multicolumn{2}{c||}{Nvidia TitanV} & \multicolumn{2}{c||}{AMD MI60} & \multicolumn{2}{c|}{Nvidia V100} \\
     \hline
     \multicolumn{1}{|c||}{Test}  & $T_0$ & $W_{\max}$ & $T_0$ & $W_{\max}$  & $T_0$  & $W_{\max}$  & $T_0$  & $W_{\max}$ \\ \hline
BS1 & 8.87 & 870  & 2.49 & 601  & 7.95 & 846  & 2.90 & 811 \\
BS2 & 9.09 & 836  & 2.71 & 620  & 8.73 & 821  & 2.88 & 846 \\
BS3 & 17.73 & 796  & 5.24 & 619  & 16.99 & 843  & 7.62 & 809 \\
BS4 & 17.62 & 747  & 5.96 & 639  & 16.69 & 831  & 7.23 & 879 \\
BS5 & 13.70 & 741  & 5.07 & 608  & 18.35 & 810  & 5.56 & 810 \\
BS6* & 12.97 & 765  & 5.32 & 595  & 15.32 & 724  & 6.00 & 769 \\
BS7* & 9.08  & 778  & 1.77 & 563  & 14.19 & 735  & 3.70 & 787 \\
    \hline
    \end{tabular}
    \caption{Summary of model fitting results for fixed kernel launch cost ($T_0$ given in $\mu{}s$) and asymptotic bandwidth ($W_{\max}$ given in GB/s). The $^*$ on BS6 and BS7 is to indicate that the model fit for these tests is for only polynomial degree $N=7$. }
    \label{BSparams.tab}
\end{table*}

\section{Summary} \label{sec:summary}

In this paper, we have proposed several new benchmark tests relevant to iterative linear solvers used in high-order finite element methods. In particular, our tests are inspired by several operations necessary in a Conjugate Gradient iterative method for hexahedral finite element meshes, as such problems are proposed and considered by the CEED co-design center \cite{ceed2017} as part of the Exascale Computing Project. We name these tests the CEED Benchmark Streaming tests, after the CEED Benchmark Problems from which they are inspired. Each of these new benchmarks is of low arithmetic intensity, and is thus ideally limited only by how quickly data can be read from and written to a compute device's main memory.

To demonstrate the utility of these benchmarks, we have implemented them using the OCCA abstraction framework \cite{medina2014occa} in order to test the performance behavior of four separate GPU accelerators from two major manufacturers, namely the Titan V and V100 GPUs manufactured by Nvidia, and the Radeon\TM VII and Radeon Instinct\TM MI60 GPUs manufactured by AMD. In our computational tests, we find that, despite their architectural differences, we are able to obtain a high percentage of the peak bandwidth of each device's HBM memory in each of our tests. We also propose a simple model for each device's performance in each test which we demonstrate very accurately captures the realized performance behavior. Fitting this performance model using the Benchmark Streaming tests helps provide a method for predicting real performance in streaming operations at different problem sizes, which is helpful for predicting behaviors of applications when scaling to many processors.

This paper has only focused on low arithmetic intensity streaming operations. In a following work we will consider the full CEED Benchmark Problems and demonstrate both high performance as well as portability between multiple GPU architectures and manufacturers.

\section*{Acknowledgments}
This research was supported in part by the Exascale Computing Project, a collaborative effort of two U.S. Department of Energy organizations (Office of Science and the National Nuclear Security Administration) responsible for the planning and preparation of a capable exascale ecosystem, including software, applications, hardware, advanced system engineering, and early testbed platforms, in support of the nation’s exascale computing imperative.

This research was also supported in part by the John K. Costain Faculty Chair in Science at Virginia Tech.

AMD, the AMD Arrow logo, Radeon, Radeon Instinct, EYPC, and combinations thereof are trademarks of Advanced Micro Devices, Inc. OpenCL is a trademark of Apple Inc. used by permission by Khronos Group, Inc. Other product names used in this publication are for identification purposes only and may be trademarks of their respective companies.

\bibliographystyle{plain}
\bibliography{PortableStreamingKernels_arxiv}

\end{document}